\documentclass[12pt,titlepage]{article}
\usepackage[a4paper,lmargin={2.5cm},rmargin={2.5cm},tmargin={2.5cm},bmargin = {2.5cm}]{geometry}

\usepackage{booktabs}
\usepackage{afterpage}
\usepackage{parskip}
\usepackage[german,english]{babel}

\usepackage{textgreek}
\usepackage[utf8]{inputenc}
\usepackage{graphicx}		
\usepackage{dcolumn}		
\usepackage{bm}				
\usepackage{amsmath}
\usepackage{amssymb}
\usepackage{mathtools}
\usepackage{siunitx}
\usepackage{acronym}
\usepackage{nicefrac}
\usepackage{csquotes}

\usepackage{subcaption}
\usepackage{wrapfig}
\usepackage[font=small,labelfont=bf]{caption}

\usepackage[backend=biber,style=numeric,autocite=plain,sorting=none]{biblatex}
\usepackage[hidelinks]{hyperref}
\hypersetup{colorlinks=true,linkcolor={blue},citecolor={blue},urlcolor={blue}}

\begin{document}

\begin{center}
{\huge \textbf{ \Large The improved inverted AlGaAs/GaAs interface: \\ its relevance for high-mobility quantum wells and hybrid systems
} }
\end{center}

E. K\"ulah\textsuperscript{1}, C. Reichl\textsuperscript{1}, J. Scharnetzky\textsuperscript{1}, L. Alt\textsuperscript{1}, W. Dietsche\textsuperscript{1} and W. Wegscheider\textsuperscript{1} \\

\textsuperscript{1} ETH Z\"urich, 8092 Z\"urich, Switzerland  
\\\\

\textbf{\Large Abstract} \\
\newline 
Two dimensional electron gases (2DEGs) realized at GaAs/AlGaAs single interfaces by molecular-beam epitaxy (MBE) reach mobilities of about 15 million $cm^2/Vs$ if the AlGaAs alloy is grown after the GaAs. Surprisingly, the mobilities may drop to a few millions for the identical but inverted AlGaAs/GaAs interface, i.e. reversed layering. Here we report on a series of inverted heterostructures with varying growth parameters including temperature, doping, and composition. Minimizing the segregation of both dopants and background impurities leads to mobilities of 13 million $cm^2/Vs$ for inverted structures. The dependence of the mobility on electron density tuned by a gate or by illumination is found to be the identical if no doping layers exist between the 2DEG and the respective gate. Otherwise, it differs significantly compared to normal interface structures. Reducing the distance of the 2DEG to the surface down to $\SI{50}{nm}$ requires an additional doping layer between 2DEG and surface in order to compensate for the surface-Schottky barrier. The suitability of such shallow inverted structures for future semiconductor-superconductor hybrid systems is discussed. Lastly, our understanding of the improved inverted interface enables us to produce optimized double-sided doped quantum wells exhibiting an electron mobility of 40 million $cm^2/Vs$ at $\SI{1}{K}$.

\newpage
\section{\textbf{\Large Introduction}} \label{intro}
Topological protected quantum computing based on Majorana fermions requires proximity induced superconductivity. Induced superconductivity has already been reported in graphene \cite{heersche2007bipolar}, topological insulators \cite{sacepe2011gate,veldhorst2012josephson} and 2DEGs in InAs both in planar \cite{takayanagi1985superconducting} and one-dimensional systems \cite{mourik2012signatures} as well as in AlGaAs/GaAs structures \cite{wan2015induced}. Here the critical current can even be controlled electrostatically. However, experimental studies on the properties of Majorana quasi particles are still scarce. Particularly, the hybrid systems, i.e. combined semiconductor-superconductor structures, are very promising candidates for investigations in this field \cite{sato2016majorana,frolov2020topological}.  \\
Unfortunately, the tunnel coupling between a 2DEG in GaAs and a superconductor is very weak because a Schottky barrier forms at the surface of the GaAs/AlGaAs heterostructure, which originates from both different work functions at the interface and from electron traps \cite{doi:10.1063/1.94599,PhysRevLett.58.1260}. This barrier is less relevant in InAs \cite{shabani2016two,takayanagi1985superconducting,lee2019transport} but GaAs is required if the the 2DEG should be in the fractional quantum-Hall state \cite{PhysRevX.4.031009, PhysRevB.93.214504, PhysRevLett.117.096803}. The standard ``MODFET" design (Fig.~\ref{fig:Structures}a) leads to an even larger barrier because an AlGaAs layer with even larger gap is placed  between the 2DEG and the interface to the superconductor. On the other hand it had been shown by Tsui already several decades ago in e.g. \cite{PhysRevLett.21.994} that tunneling between a superconductor into highly doped GaAs is possible. \\
In this publication, we investigate properties of the  AlGaAs/GaAs ``inverted" structure (see Fig.~\ref{fig:Structures}b). One motivation compared to a MODFET device is the possibility to shift the wave function of the electrons closer to a superconductor deposited on the semiconductor surface. The reason for this is the absence of the high barrier AlGaAs towards the surface-facing side of the interface. Therefore, a better coupling of the 2D electrons and the Cooper pairs in case of an adjacent superconductor is expected as opposed to the use of a MODFET. In addition, the coupling strength is further increased with smaller 2DEG-to-surface distances. \\ 
Apart from Majorana physics, the inverted structure could be more beneficial than the MODFET design with respect to top gate stability. This is expected due to the same reason that neither dopants nor AlGaAs exist between the 2DEG and the surface. This would eliminate the so called ``DX-centers" (deep donors with high ionization energies \cite{mooney1990deep}) that occur only in doped AlGaAs and are known to cause gating instabilities. \\
Traditionally, the inverted interface is said to have inferior quality compared to the well-established MODFET \cite{rauch1991scattering,kohrbruck1990inequivalence, saku1996limit}. Despite improved inverted 2DEG mobilities by applying graded superlattices at the interface \cite{fischer1984improvement,sajoto1989use}, increased setback layers \cite{pfeiffer1991si}, top- \cite{kane1995high} and back gating \cite{saku1998high}, this hierarchical order remained unchanged. Therefore, it is of interest to investigate the inverted interface heterostructures. Here, scattering is of particular importance. Additional knowledges will in turn be profitable in other high(est)-mobility structures, i.e. double-sided doped quantum wells (DSQWs) (Fig.~\ref{fig:Structures}c) which contain also an inverted interface. \\
In summary, we employ the technique of molecular beam epitaxy (MBE) to synthesize heterostructures of extraordinary material quality. We focus especially on the growth of inverted AlGaAs/GaAs single interface structures. For the first time, 2DEGs with comparable electron mobility to the well established MODFETs are realized. In addition, we set out to minimize the distance from the 2DEG to the surface. Top-and back gates are fabricated as well to these heterostructures. Finally, adding top-and back gates to DSQWs lead to 2DEGs with remarkable high mobilities by centering the wave function in the QW. DSQWs with highest quality are an indispensable prerequisite to explore the nature of certain exotic fractional quantum Hall states, for example the non-Abelian state at filling factor $\nu = 5/2$. \\


\section{\textbf{\Large Experimental details}} \label{experimental_details}
\begin{figure}[h]
	\centering
	\includegraphics[width=0.6\textwidth]{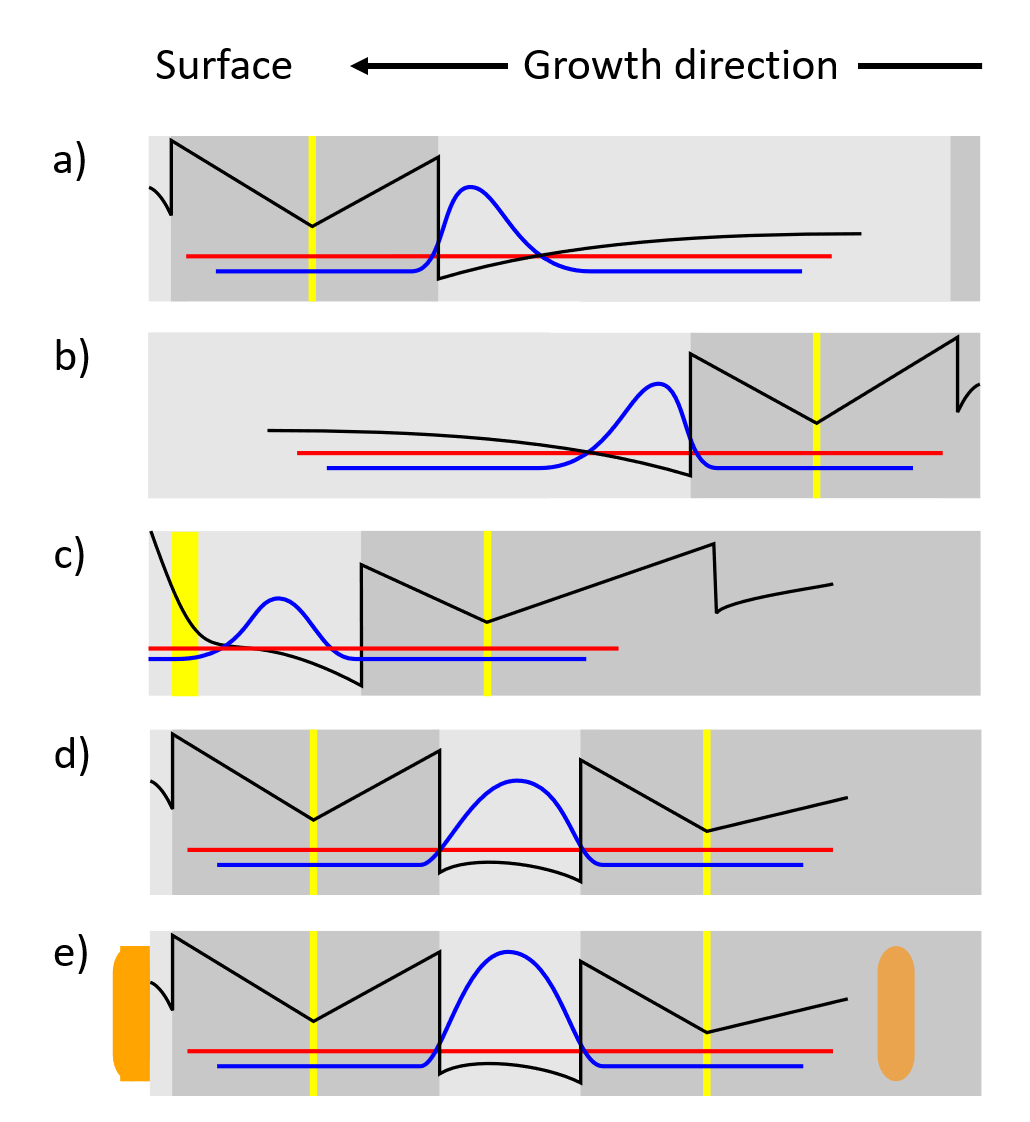}
	\caption{The growth sequences with corresponding conduction bands (Fermi level in red) as well as ground state electron wave functions (blue) are given for a) a MODFET, b) an inverted structure, c) a shallow inverted structure, d) a DSQW, and e) a top-and back-gated DSQW, respectively. Bright/dark gray areas indicate GaAs/AlGaAs, respectively, and Si dopant regions are indicated in yellow. The orange areas in e) illustrate the top and the (patterned) back gate.}
	\label{fig:Structures}
\end{figure}

All samples discussed in this work are grown in a modified Veeco Gen II molecular beam epitaxy (MBE) system. The native substrate oxide is thermally desorbed in-situ, and growth rate calibrations are performed daily with reflection high-energy electron diffraction (RHEED) to ensure high precision of layer thicknesses and Al-mol fractions. Growth temperatures are around $630 \si{\degree}$C and total deposition rates of $ \sim \SI{2.6}{\angstrom/s}$ are used. \\
Schematics of the investigated heterostructures are shown in Fig.~\ref{fig:Structures}. All structures contain a buffer stack/superlattice structure (SLS) of AlAs/GaAs to reduce surface roughness and to getter residual impurities in the main vacuum chamber. \\
The ``standard" structure consists of a single GaAs/AlGaAs heterointerface with a thin layer of silicon dopants residing between the interface and the sample surface, henceforth referred to as MODFET (``modulation doped single interface", Fig.~\ref{fig:Structures}a). We use the term of MODFET throughout the whole work although it is not compulsory a FET. In case of an additional gate application, it is then appropriate written. \\
``Inverted" structures consist of a doping layer in AlGaAs, capped by a GaAs layer hosting the 2DEG (Fig.~\ref{fig:Structures}b). The thickness of the capping layer is varied from $\SI{1000}{nm}$ down to $\SI{50}{nm}$. For the latter ``shallow" structures, an additional doping layer (either Si-$\delta$ or Si-GaAs) is inserted to provide sufficient conduction band bending (Fig.~\ref{fig:Structures}c). This turned out to be necessary in order to prevent 2DEG depletion resulting from Fermi level pinning in mid-gap position at the GaAs surface. \\
Combining the MODFET and the inverted structure results in a symmetric quantum well of GaAs with thin doping layers on both sides (double-sided doped quantum well, DSQW, Fig.~\ref{fig:Structures}d). Setback distances are in the range of about $\SI{30}{nm}$ to $\SI{100}{nm}$. The Al mol-fraction is set to $\SI{30}{\%}$ throughout most of the structures, but reduced to $\SI{24}{\%}$ in the regions close to the 2DEG. Electrical gating for the inverted 2DEGs is realized by placing a highly Si-doped GaAs layer underneath the initial buffer layer stack (back gates) or by ex-situ Al evaporation on the surface (top gates). More advanced so-called patterned back gates \cite{berl2016structured} are applied for the DSQWs (Fig.~\ref{fig:Structures}e). \\
Magnetotransport measurements are performed using standard lock-in techniques on Van der Pauw samples as well as using Hallbar geometries. 
The experimental data discussed in this work was obtained mainly in an evacuable $He4$ dipstick cryostat or using a dilution refrigerator, at electron temperatures of $\SI{1.3}{K}$ Kelvin and $< \SI{60}{mK}$ respectively.  \\

\newpage
\section{\textbf{\Large Results}} 

\subsection{Improved growth of the inverted 2DEG and its quality on par with the MODFET} \label{benchmark_inverted}
\begin{figure}[h]
	\centering
	\includegraphics[width=1\textwidth]{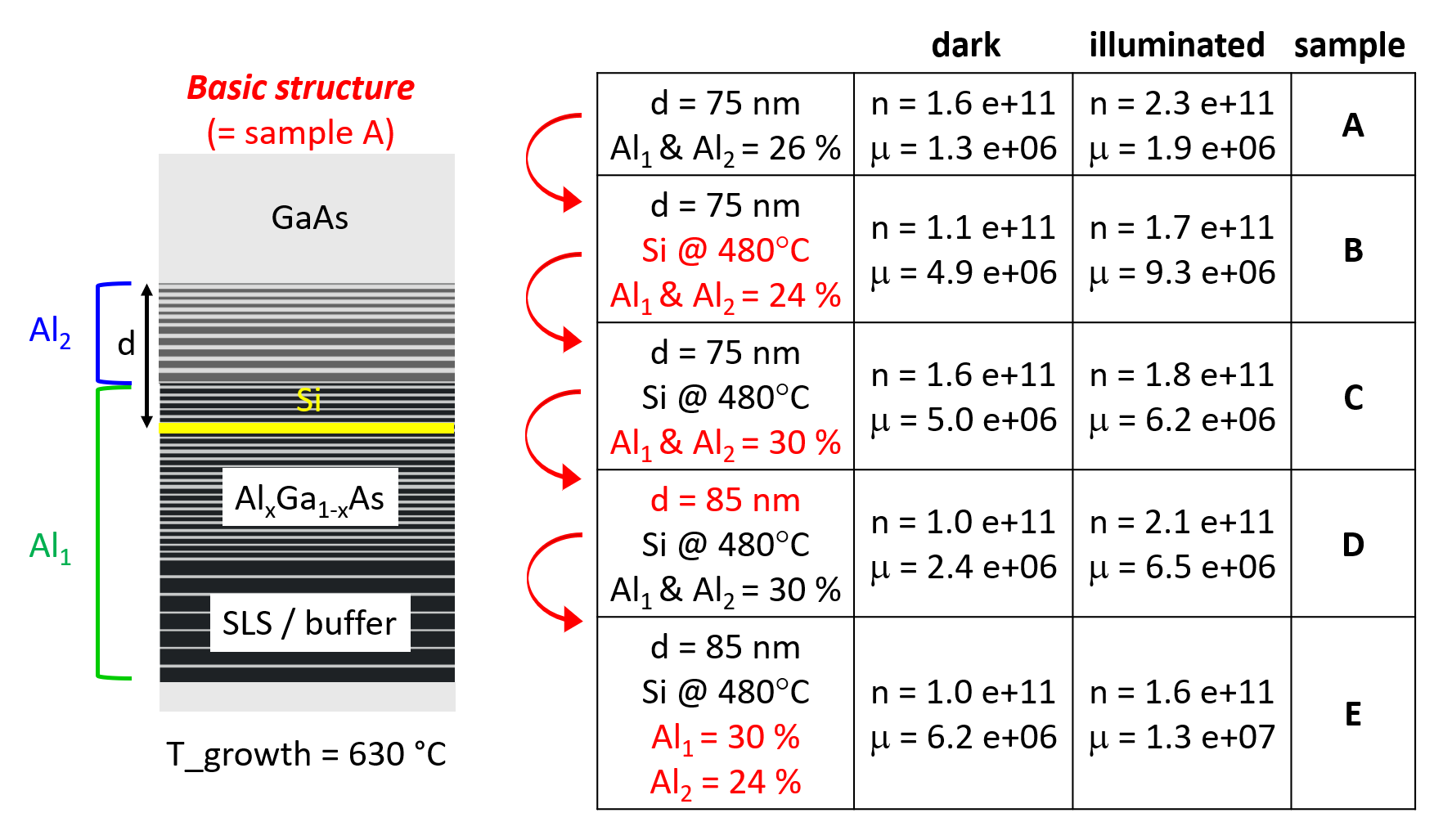}
	\caption{Left: Our basic inverted interface at the beginning of our study towards high-mobility structures. Right: Modified structural compositions with corresponding mobilities/densities are listed. The mobility is given in terms of $cm^{2}/Vs$, the electron density in $cm^{-2}$ and all values were recorded at $\SI{1}{K}$.}
	\label{fig:Inverted_benchmark_changes}
\end{figure}

It is well established that the inverted interface is inferior in 2DEG quality compared to the MODFET \cite{saku1994high,sasa1984improved,morkoc1982interfacial}. This is true even if identical growth procedures are applied to both structures, just in reversed order. The highest mobility of an inverted 2DEG reported in literature is $\mu = 5\cdot 10^{6}{cm^2/Vs}$ \cite{saku1998high} with corresponding MODFETs of $\mu = 10\cdot 10^{6}{cm^2/Vs}$ \cite{saku1996limit}. \\
Several processes have been identified that could be detrimental for the inverted interface. In this publication we discuss three different scattering sources, namely: (i) remote ionized impurity (RI) scattering from the doped $\delta$-layer in the AlGaAs, (ii) background impurity (BI) scattering as a result of incorporated residuals from the growth environment, and (iii) interface roughness (IR) scattering that comes from the segregation of any dopant or ``gettered" impurity plus the intermixing of the Al atoms. In addition, the dominants of the IR scattering strongly depends on the penetration of the wave length into the barrier. \\
The structural quality will be evaluated via the low-temperature mobility (density) of the 2D electrons. We assume that the mobility exhibits a density dependence $\mu \propto (n_{2DEG})^{\alpha}$ where the $\alpha$\, is characteristic for the dominant scattering mechanism. Sample illumination with a red LED at low temperature is applied, which releases additional carriers from DX centers inducing increased 2DEG densities. \\ 
We start with a basic inverted heterostructure that consists of a $\sim \SI{1.5}{\mu m}$ buffer stack, a $\SI{75}{nm}$ setback layer, a single-sided Si QW-doping \cite{baba1986elimination} (i.e. the Si dopants located in a $\sim \SI{1}{nm}$-wide GaAs QW in the AlGaAs), an Al mol-fraction of $A_{1} = A_{2} = \SI{26}{\%}$ throughout the whole structure and distances of $\sim \SI{500}{nm}$ to the GaAs surface (Fig.~\ref{fig:Inverted_benchmark_changes} left). A typical short-period superlattice structure (SLS) is used instead of a homogeneous AlGaAs layer. This is quite well-known and was already employed by others \cite{umansky2009mbe} where alternating Al(Ga)As and GaAs layers capture BIs in the multiple QWs in growth direction. In addition, such a SLS serves the purpose of smoothing the surface leading to lower IR at the 2DEG region. Each GaAs layer is of $\SI{1.132}{nm}$\, thickness except after the Si doping layer, where it is reduced to $\SI{0.566}{nm}$. The AlAs layers vary from [$4, 3, 2.264, 1.132$] $nm$ in the SLS, and the $Al_{x}Ga_{1-x}As$ includes [$8, 22, 9.434, 4.528$\,] $nm$\, in direction to the interface. A constant growth temperature of $630 \si{\degree}$C is employed throughout the whole structure. The mobility of these 2DEGs is only $\mu=1.3 \cdot 10^{6}{cm^2/Vs}$ at $n = 1.6 \cdot 10^{11}{cm^{-2}}$ for the dark state, and $\mu = 1.9 \cdot 10^{6}{cm^2/Vs}$ at $ n = 2.3 \cdot 10^{11}{cm^{-2}}$ for the illuminated state, respectively (Fig.~\ref{fig:Inverted_benchmark_changes} tabular, sample A). \\
It is known that dopant migration \cite{jansen1990migration, inoue1985effects,gonzalez1986silicon} occurs towards the inverted interface during the growth. This increases not only the relevance of RI scattering, but additionally makes it much harder to predict and control the electrostatic effect of any doping layer in the structure. Therefore, in a next step we reduce the substrate temperature only during the Si doping procedure from normally $630 \si{\degree}$C to $480 \si{\degree}$C (Fig.~\ref{fig:Inverted_benchmark_changes} tabular, sample B). We speculate the dopants to be frozen in the growing crystal and that their migration is avoided. A clear improvement in the 2DEG quality is visible: $\mu = 5 \cdot 10^{6}{cm^2/Vs}$ ($n = 1.1 \cdot 10^{11}{cm^{-2}}$) for the dark state and $\mu = 9.3 \cdot 10^{6}{cm^2/Vs}$ ($n = 1.7 \cdot 10^{11}{cm^{-2}}$) for the illuminated state, respectively. We conclude that Si dopant segregation along the growth direction is drastically minimized. Moreover, we would like to point out that even completely undoped structures, with the inverted 2DEG induced only by a back gate, exhibit a similar maximum mobility of $\mu = 5 \cdot 10^{6}{cm^2/Vs}$ \cite{saku1998high}.\\
In addition to dopant migration, the segregation of the Al atoms can appear as well \cite{moison1989surface}, which widely expands into the AlGaAs especially for lower Al concentrations. This provokes crystal disorder and alloy scattering. In contrast, the Al segregation into the GaAs is absent for the MODFET since the AlGaAs follows the interface. Related to that, we studied the influence of thin (``delta'') Al layers directly in regions the electrons reside for a MODFET \cite{reichl2015mapping}. The outcome was that mobilities suffered substantially according enhanced electron scattering rates. In addition, the 2DEG deteriorates further due to residual impurity incorporation, caused by the highly reactive Al atoms. \\
Based upon these observations, we changed in the second step the Al concentration in the $Al_{x}Ga_{1-x}As$ to investigate the quantitative impact of either higher or lower alloy contents for the inverted structure. The Al-mol fraction is increased from $A_{1} = A_{2} = \SI{24}{\%}$ to $A_{1} = A_{2} = \SI{30}{\%}$ (Fig.~\ref{fig:Inverted_benchmark_changes} tabular, sample C). The mobility is again $\mu = 5 \cdot 10^{6}{cm^2/Vs}$ in the dark state, however, with a $\SI{45.5}{\%}$ higher density than in the previous sample B. Usually, one would expect a linear behavior of $\mu$ related to $n$, which is not the case from sample B to sample C. This confirms the stronger IR and BI scattering as a result of the higher Al concentration in the AlGaAs sequence. Note also that the higher density in sample C (dark state) comes from possibly increased amount of DX centers \cite{mooney1990deep}. In total, the inverted interface is worse in quality. The same is true after illumination with a mobility drop of $\mu =3 \cdot 10^{6}{cm^2/Vs}$ compared to sample B for almost identical densities. \\
So far, we discussed the segregation of Si and Al during the growth process. Interestingly Pfeiffer at al. achieved enhanced mobilities by using larger setback distances in their inverted interfaces \cite{pfeiffer1991si}. In doing so, the scattering of migrated dopants is kept more far from the crucial interface. Likewise, we increased in our third step of growth modification the setback layer width from $\SI{75}{nm}$ to $\SI{85}{nm}$ (Fig.~\ref{fig:Inverted_benchmark_changes} tabular, sample D). No profound change is evident except a lower mobility in the dark state, however, also at lower density. We ascribe this to the lower dopant excitation probability into the 2DEG because of an increased setback distance. Only illumination enables the charge carriers to overcome the potential energy barrier, which is verified in the illuminated result. Consequently, the substrate temperature down-ramping solves already the migration and there is no need of an additional increased setback width. In contrast, Pfeiffer et al. applied a constant temperature throughout their whole structure \cite{pfeiffer1991si}. \\
In our last change, we focus on BI scattering. The idea is to collect and keep away residual impurities as much as possible from the inverted interface. Therefore, different Al concentrations are employed for different layer sequences along the growth direction. In detail, a higher Al content is used first at the beginning for the buffer stack and then lowered in the setback layer. We use $ Al_{1} = \SI{30}{\%}$ and $ Al_{2} = \SI{24}{\%}$ for sample E (Fig.~\ref{fig:Inverted_benchmark_changes} tabular). This leads to our final benchmark for an inverted 2DEG with $\mu = 6.2 \cdot 10^{6}{cm^2/Vs}$ at $n = 1.0 \cdot 10^{11}{cm^{-2}}$, and $\mu = 13 \cdot 10^{6}{cm^2/Vs}$ at $n = 1.6 \cdot 10^{11}{cm^{-2}}$ in the dark and illuminated state, respectively. \\ 

\begin{figure}[h]
	\centering
	\includegraphics[width=0.75\textwidth]{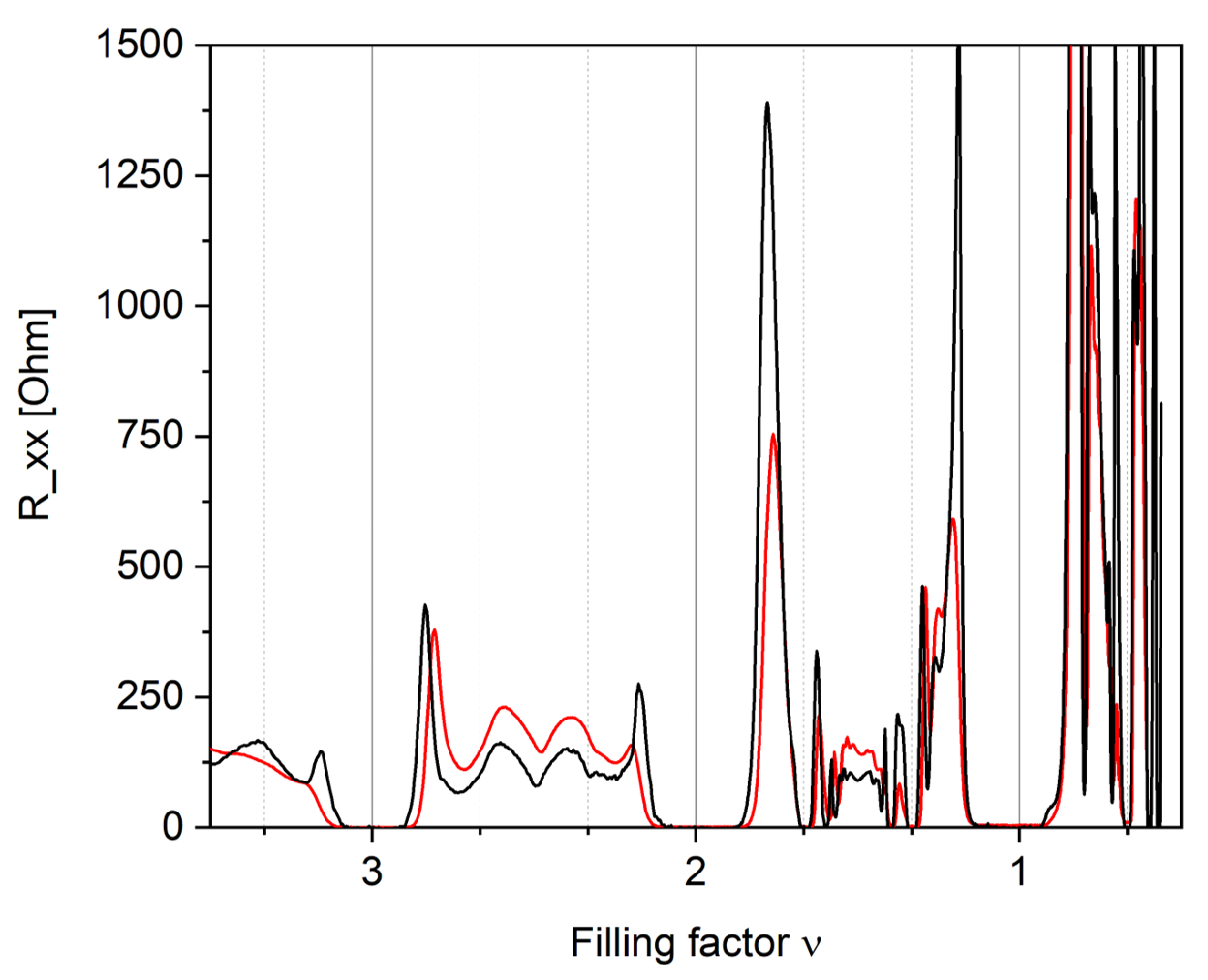}
	\caption{Comparing magnetotransport data of the benchmark MODFET and the benchmark inverted structure at $\SI{60}{mK}$ after low-temperature illumination. The black trace denotes the inverted 2DEG and the red one the MODFET, respectively.}
	\label{fig:Comparison_Inverted_MODFET}
\end{figure}

Our $\alpha$ exponents derived from the relation $\mu \propto (n_{2DEG})^{\alpha}$ are between $0.6$ and $0.64$ for the results in Fig.~\ref{fig:Inverted_benchmark_changes} (only two density/mobility values are used for each sample, dark and illuminated state). Hence, BI scattering represents the dominant scattering mechanism for this type of inverted heterostructures. This is in agreement with other works examining inverted 2DEGs \cite{pfeiffer1991si, saku1998high}. \\
If we now compare our benchmark-inverted-2DEG with our standard MODFETs' benchmark ($\mu = 15 \cdot 10^{6}{cm^{2}/Vs}$,  $n = 1.8 \cdot 10^{11}{cm^{-2}}$) \cite{reichl2014mbe}, extrapolation of the inverted density to $n = 1.8 \cdot 10^{11}{cm^{-2}}$ also gives $\mu = 15 \cdot 10^{6}{cm^{2}/Vs}$ assuming a mobility dependence $\mu \propto (n_{2DEG})^{\alpha}$ with $\alpha \approx 0.62$. \\
A more profound analysis of the sample quality is obtained by means of magnetotransport spectroscopy. We compare the longitudinal and the Hall resistance of the benchmark MODFET with that of the benchmark inverted structure at a temperature of $\SI{60}{mK}$. The overall features are in good qualitative agreement and thus, underline the comparable quality of the two sample types (Fig.~\ref{fig:Comparison_Inverted_MODFET}). \\

We would like to point out the fact that other types of heterostructures also profit from the improved inverted interface. This is particularly true for our highest-mobility 2DEG in a DSQW (chapter~\ref{40mio}). \\

\subsection{Shallow inverted AlGaAs/GaAs heterostructures} \label{shallow_inverted}
\begin{figure}[h]
	\centering
	\includegraphics[width=1\textwidth]{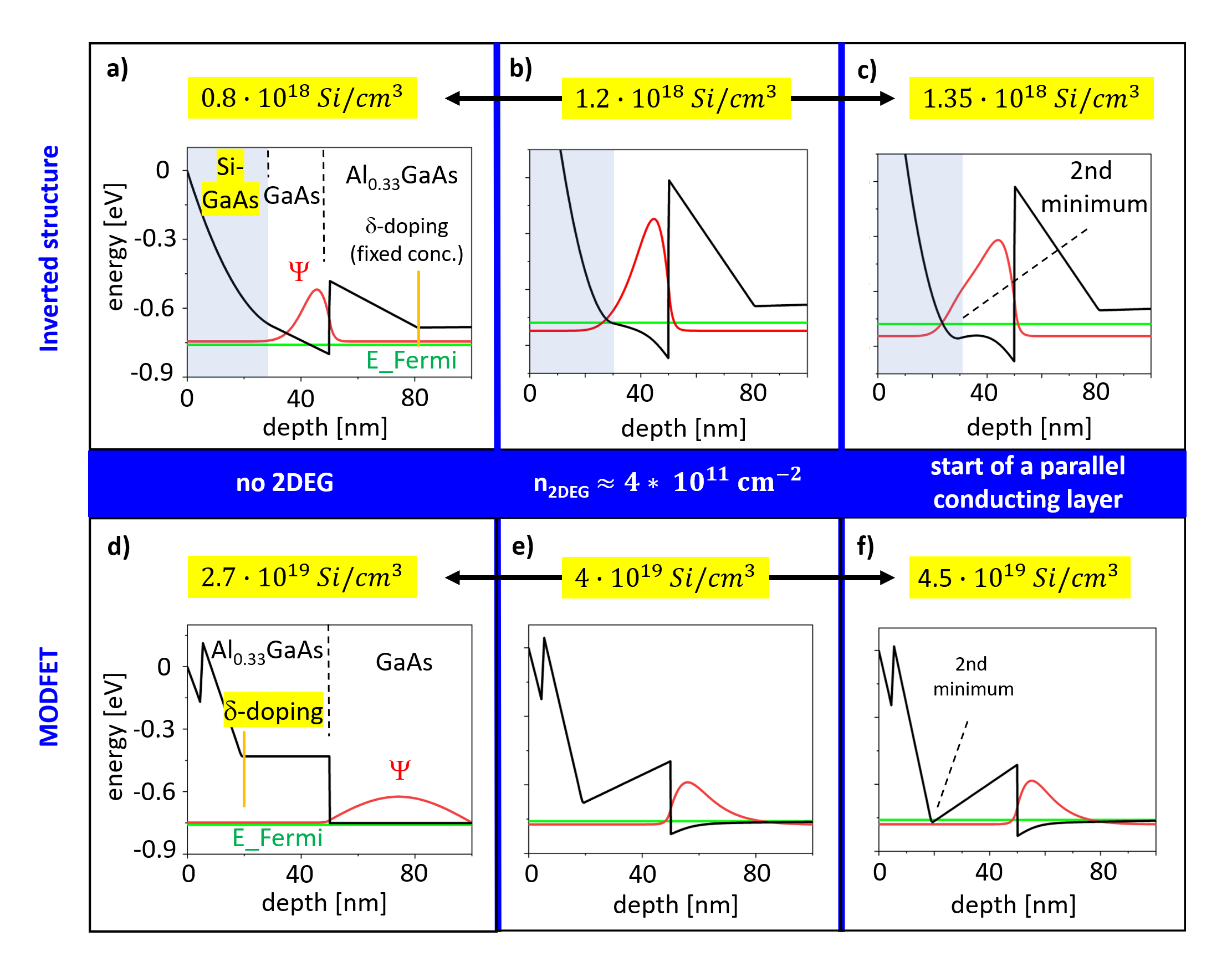}
	\caption{One-dimensional simulations are done on the basis of shallow inverted structures (a to c) and MODFETs (d to f) using the Schr\"odinger-Poisson solver AQUILA software. The critical amount of the Si doping concentration to the surface-facing side is investigated for a $\SI{50}{nm}$ deep 2DEG. In case of the inverted structure, only the amount of dopants in the $\SI{30}{nm}$ thick Si-GaAs layer was varied while keeping the Si $\delta$-doping in the AlGaAs constant. For the MODFET, the Si $\delta$-doping was modified with only one doped layer throughout the whole structure. The electron wave function $\Psi$ of the ground state is given in red, the Fermi energy level in green and the conduction band in black, respectively. A 2DEG with a density of $n \approx 4 \cdot 10^{11}{cm^{-2}}$ is formed as illustrated in b) and e), and the impact of either lower (a and d) or higher (c and f) Si concentrations is shown.}
	\label{fig:Simulation}
\end{figure}

Present quantum matter research is often based upon hybrid systems, which refers to the interaction between quantum states of different materials. As already introduced at the beginning of this publication, the semiconductor-superconductor combination is a prime example. A prerequisite is an interface, where Cooper pairs are able to transfer and diffuse into the semiconductor due to the proximity effect. It is known that no Schottky barrier is formed in InAs at the surface. Here, the surface Fermi level is located above the conduction band minimum and thus, a so called inversion layer occurs. On the other hand, 2DEG qualities are much worse when compared to Al(Ga)As/GaAs systems. This drawback is a serious obstacle in proximity experiments where more fragile and exotic quantum states can occur only in 2DEGs of higher quality. \\
In the following, which is a simulation section, we discuss if our inverted interfaces offer advantages over MODFETs for producing hybrid systems. We focus on shallow structures whose 2DEGs should have minimum distances to an interfacing superconductor. This is an interesting alternative approach compared to other techniques, e.g. metal/alloy depositions on top of the semiconductor, or a chemical surface treatment in order to lower the Schottky barrier height at the surface. Emphasis must be on a sizable tunnel probability of the Cooper pairs into the 2DEG. Our shallow inverted structures offer the possibility to shift the electron wave functions towards the surface for such a higher coupling strength. One possibility would be top gating with the superconductor used directly as a gate, another one additional doping. \\  
Our shallow inverted heterostructures have 2DEG depths between $\SI{100}{nm}$ and $\SI{50}{nm}$. It turned out that a second doping in the GaAs is inevitable to create a 2DEG (Fig.~\ref{fig:Structures}c). The Schottky barrier of GaAs is quite high \cite{myburg1998summary}, hence, an adequate conduction band bending by doping is necessary. The ionization energy of the doping atoms is smaller in pure GaAs than in AlGaAs. This leads to a larger effective Bohr radius and to a stronger overlap of neighboring donor electrons in GaAs. As a consequence, parallel conductance occurs already at lower doping levels than in AlGaAs. This makes the growth of shallow inverted structures a challenge. Indeed, the success rate of such structures has been only about $\SI{20}{\%}$ compared to nearly $\SI{100}{\%}$ for deep inverted ones. \\
The amount of Si in the $\delta$-doping layer is kept constant. These shallow 2DEGs are much harder to synthesize in contrast to ``deep" inverted structures as discussed in chapter~\ref{benchmark_inverted} (Fig.~\ref{fig:Structures}b). The reason is that the doping level in the Si-GaAs layer must be adjusted very precisely. It should not form a parallel conducting layer, but being efficient enough to form the 2DEG. \\
\newline
Fig.~\ref{fig:Simulation} shows different simulation scenarios for a $\SI{50}{nm}$ deep 2DEG of an inverted structure and a MODFET, respectively. Identical growth settings are adjusted, i.e. the Al-mol fraction, 2DEG density and interface depth. The Si doping concentration between the interface and the surface ($\SI{0}{nm}$) is investigated with respect to the electrons in their ground state. The conduction band is illustrated in black, the electron wave function $\Psi$ in red and the Fermi energy level in green, respectively. The additional Si-doped GaAs in the inverted case is of $\SI{30}{nm}$ thickness, which is represented by the blue square area from a) to c). This is not needed for the MODFET where surface states are compensated by the $\delta$-doping layer in the AlGaAs. \\
In the simulations in Fig.~\ref{fig:Simulation}b) and Fig.~\ref{fig:Simulation}e), the 2DEG exists in both structures at the interface ($\SI{50}{nm}$) with equal densities of $n \approx 4 \cdot 10^{11}{cm^{-2}}$. If we now lower the Si concentration from $1.2 \cdot 10^{18}$ $Si/cm^{3}$ to $0.8 \cdot 10^{18}$ $Si/cm^{3}$ in the Si-GaAs for the inverted structure (Fig.~\ref{fig:Simulation}a) and from $4 \cdot 10^{19}$ $Si/cm^{3}$ to $2.7 \cdot 10^{19}$ $Si/cm^{3}$ in the $\delta$-doping layer for the MODFET (Fig.~\ref{fig:Simulation}d), then the wave function moves above the Fermi level energy, preventing the formation of a parallel channel. On the other hand, if we increase the Si amount from the cases in Fig.~\ref{fig:Simulation}b) to $1.35 \cdot 10^{18}$ $Si/cm^{3}$ and Fig.~\ref{fig:Simulation}e) to $4.5 \cdot 10^{19}$ $Si/cm^{3}$, then a parallel conducting layer starts to form (denoted as $2$nd energy minimum in Fig.~\ref{fig:Simulation}c) and Fig.~\ref{fig:Simulation}f), respectively). \\
In conclusion, both heterostructures react identical sensitive to the doping concentration at the surface-facing side before no conductivity ($\sim \SI{33}{\%}$ difference in $Si/cm^{3}$) or a parallel conducting layer ($\sim \SI{13}{\%}$ difference in $Si/cm^{3}$) occurs. The required doping precision is at the limit concerning the chamber and material state, i.e. typical variations using daily calibration tools. \\

\begin{figure}[h]
	\centering
	\includegraphics[width=0.65\textwidth]{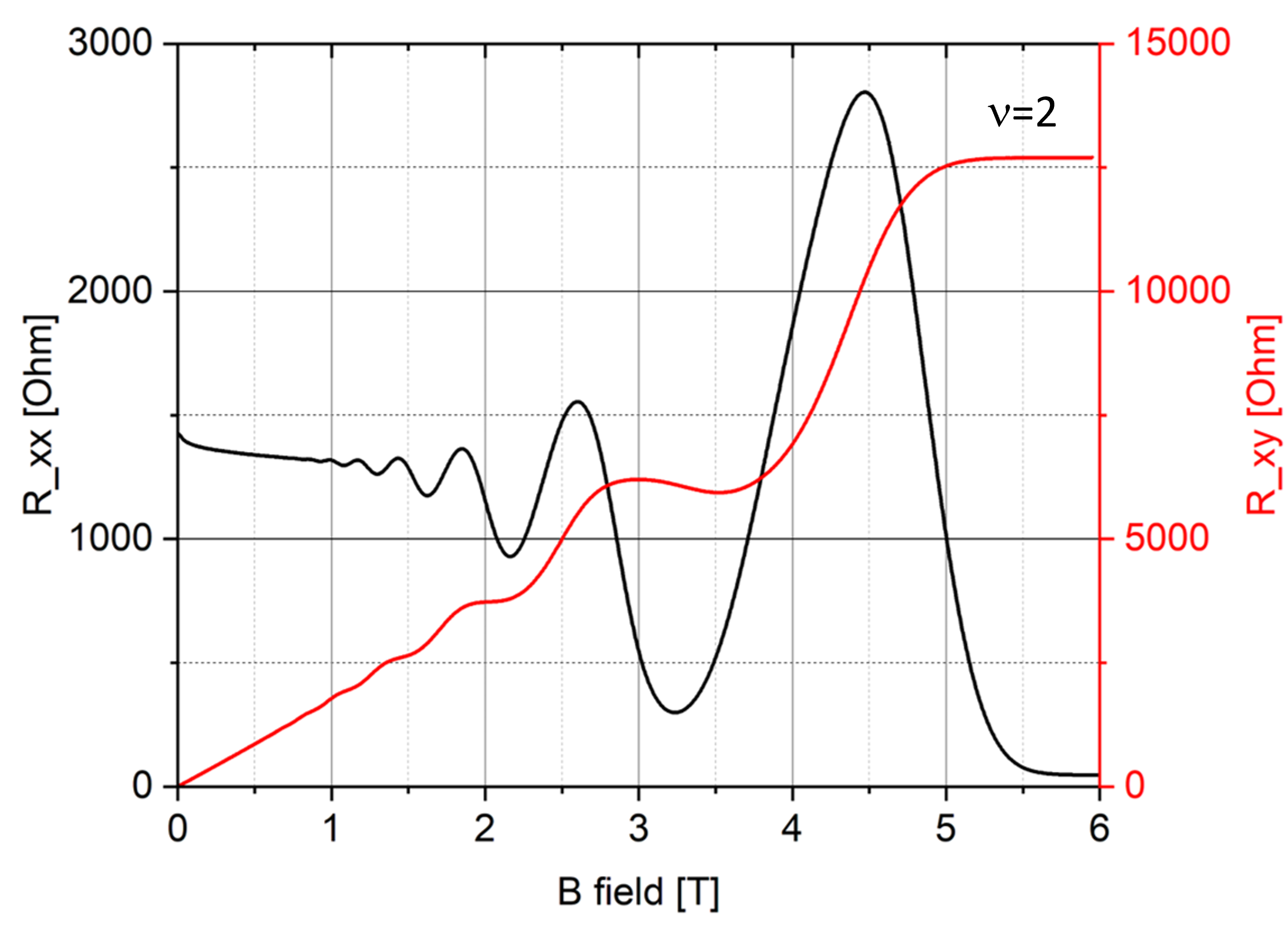}
	\caption{Longitudinal (black) and Hall (red) resistances for a $\SI{50}{nm}$ shallow inverted structure at $\SI{1}{K}$, with $\mu = 9 \cdot 10^{3}{cm^2/Vs}$ at $n = 4 \cdot 10^{11}{cm^{-2}}$.}
	\label{fig:Shallow_Inverted}
\end{figure}

Moreover, simulations show that the tolerance window of the doping deviation becomes narrower with the 2DEG getting closer to the surface. While the structures in Fig.~\ref{fig:Simulation}a-c) and Fig.~\ref{fig:Simulation}d-f) reveal a total tolerance window of $\sim \SI{46}{\%}$, this number is increased to $\sim \SI{61}{\%}$ and $\sim \SI{72}{\%}$ for identical simulations including 2DEG depth of $\SI{70}{nm}$ and $\SI{100}{nm}$, respectively. This could be a serious problem in the realization of extreme shallow structures and a challenge in growth. Additionally, steadily increased Si concentrations are found to be required by approaching the surface. However, the highest achievable doping concentrations in GaAs are lower than those in AlGaAs in case of ``standard" growth temperatures above $\sim 600 \si{\degree}$C \cite{newman1999upper,ketterer2010compensation}. Therefore, the feasibility of especially extreme shallow inverted 2DEGs is even more difficult. In contrast, MODFETs were synthesized with channel depth down to $\SI{34}{nm}$ \cite{simonet2017hybrid, tseng2011tip}. \\
Despite such circumstances, the inverted interface benefits from the absence of AlGaAs as well as of the DX-centers in doped-AlGaAs at the surface-facing side with respect to hybrid systems. Moreover, its electron wave function distribution is more favorable compared to the one of the MODFET (Fig.~\ref{fig:Simulation} a-f) because it extends closer to the surface. \\
\newline
In addition, we calculate the electron wave function probability $P = \int_{0}^{d} \bigl| \Psi \bigl|^2 dz $ for the cases in Fig.~\ref{fig:Simulation}b-c), e-f). In case of Fig.~\ref{fig:Simulation}b), a value $P = 0.0013$ is derived for the shallow inverted structure in the range between $\SI{0}{nm}$\, and $\SI{30}{nm}$\,. It is further increased to $P = 0.0053$ in Fig.~\ref{fig:Simulation}c). This probability is substantially reduced for the MODFET structure to only $5 \cdot 10^{-5}$ and $8 \cdot 10^{-5}$ for Fig.~\ref{fig:Simulation}e) and f), respectively (in the range from $\SI{0}{nm}$ to $\SI{48}{nm}$). Hence the inverted interface is indeed superior in terms of our coupling intentions. \\
\newline 
To conclude this section, we managed to successfully grow an inverted 2DEG located only $\SI{50}{nm}$ underneath the surface. We could simulate the structure with an electron density of $n \approx 4 \cdot 10^{11}{cm^{-2}}$ (Fig.~\ref{fig:Simulation}b). Its well-pronounced Shubnikov-de Haas and quantum Hall features are shown in Fig.~\ref{fig:Shallow_Inverted}. The slope of the longitudinal resistance $R_{xx}$ is nearly constant and only slight negative at low magnetic fields, hence not referring to a weak localization which usually includes a quadratic decrease before the onset of the Shubnikov-de Haas oscillations. Moreover, its minimum at filling factor $\nu = 2$ does not drop completely to zero as well as the corresponding Hall resistance $R_{xy}$ being unequal of $\SI{12.9}{k \Omega}$. We attribute all these characteristics to the very low mobility of $\mu = 9 \cdot 10^{3}{cm^2/Vs}$. This in turn is assumed as a result of the additional dopants from the Si-GaAs layer as well as strong surface state effects. It is also possible that the electron wave function is broadened towards the surface into regions of doped-GaAs. For comparison, the mobility of an inverted 2DEG with $\SI{100}{nm}$ depth reached $\mu = 3.5 \cdot 10^{6}{cm^2/Vs}$. \\

\subsection{Back gating of inverted structures} \label{backgate}
\begin{figure}[h]
	\centering
	\includegraphics[width=1\textwidth]{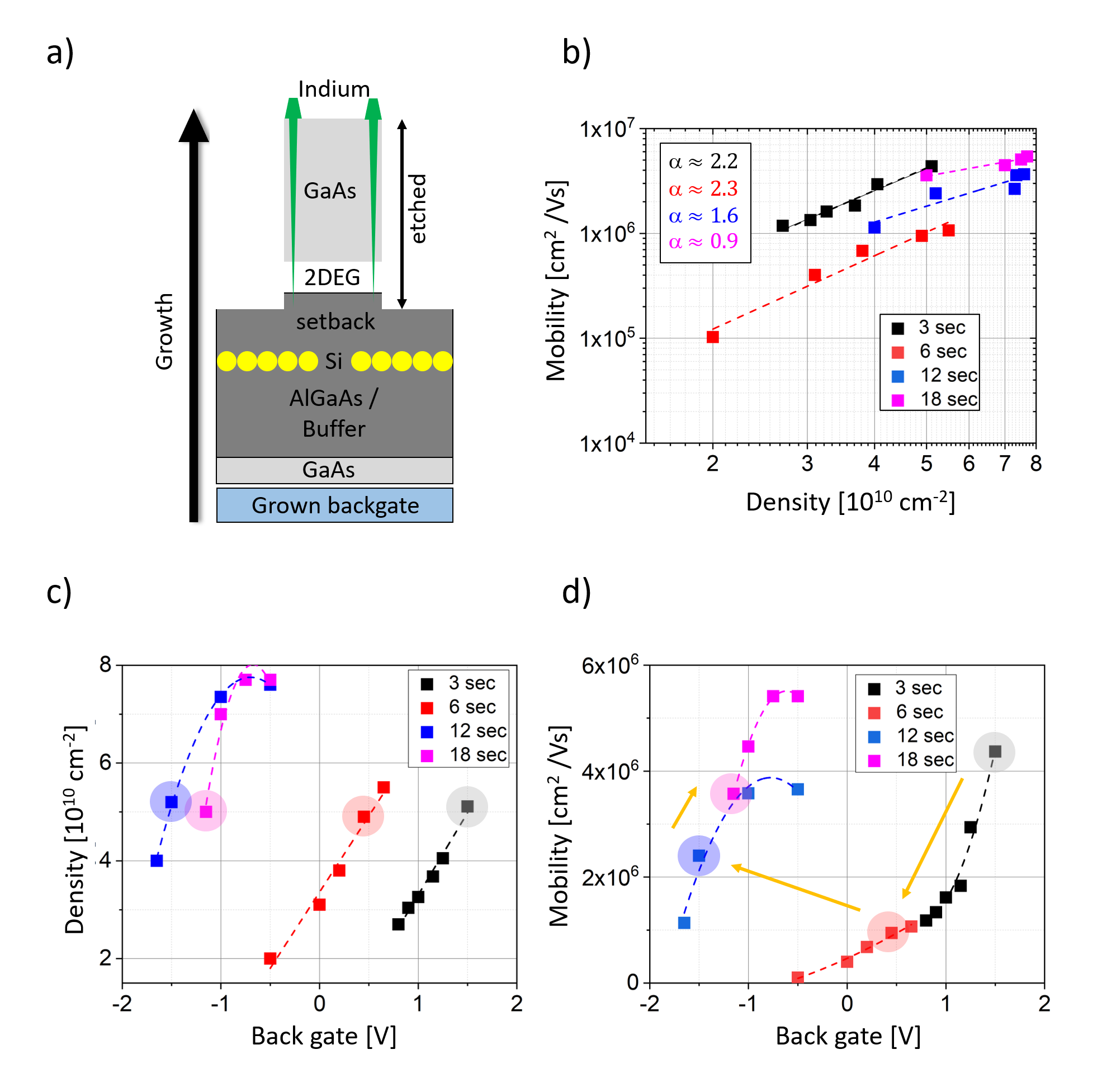}
	\caption{a) Cross-section of the back-gated inverted structure with a 2DEG tuned by a global back gate. Indium contacts were annealed from top to the interface on a Hallbar structure. The amount of Si dopants in the $\delta$-doped layer was varied for the different samples. b) Corresponding electron mobility over density in logarithmic scale. Inset: deduced $\alpha$ value. c) The electron density as a function of the global back gate voltage and d) the electron mobility over the back gate voltage. The circled dots represent the 2DEG at $n = 5 \cdot 10^{10}{cm^{-2}}$.}
	\label{fig:Backgate_Inverted}
\end{figure}

In chapter~\ref{benchmark_inverted}, different growth characteristics are discussed and highlighted for the inverted 2DEG in order to improve electron mobility. In the following, we want to understand better the impact of the $\delta$-doping layer (Fig.~\ref{fig:Structures}b), and how the interface quality is influenced by the dopant concentration.  \\
We focus on ``deep" inverted 2DEGs with the interface at least $> \SI{500}{nm}$ below the surface. The use of an additional back gate enables us to change the 2DEG density for analyzing the Si influence (Fig.~\ref{fig:Backgate_Inverted}a). The structure contains of a $\SI{65}{nm}$ setback. A growth temperature of $630 \si{\degree}$C including a down-ramping only during the doping and an Al content of $\SI{24}{\%}$ is used throughout the whole structure. This slightly differs from our highest-mobility inverted 2DEG from chapter~\ref{benchmark_inverted}. \\ 
Identical growth conditions are used for each sample. Only the doping times are varied.  \\
We use $\SI{3}{s}$ ($\equiv$ sheet density $ = 5.1 \cdot 10^{10}{cm^{-2}}$), $\SI{6}{s}$ ($\equiv$ sheet density $= 1 \cdot 10^{11}{cm^{-2}}$), $\SI{12}{s}$ ($\equiv$ sheet density $= 2 \cdot 10^{11}{cm^{-2}}$) and $\SI{18}{s}$ ($\equiv$ sheet density $= 3.1 \cdot 10^{11}{cm^{-2}}$). No conductivity is observed for the $\SI{3}{s}$\, doped sample at zero gate voltage. In comparison, a 2DEG exists for $\SI{6}{s}$ with a density of $n = 3 \cdot 10^{10}{cm^{-2}}$. The $\SI{12}{s}$ and $\SI{18}{s}$ samples exhibit a parallel conducting layer directly from the beginning (also at zero gate voltage). It is important to note that no leakage to the back gate occurs after annealing the Indium contacts from the surface to the 2DEG (Fig.~\ref{fig:Backgate_Inverted}a). It is also quite remarkable that the $\SI{3}{s}$ doped structure reveals a 2DEG after gating although nearly no dopants are available. In such a device resembling a heterostructure isolated gate field effect transistor (HIGFET) structure, the 2D channel is formed predominantly by the electric field. \\
Fig.~\ref{fig:Backgate_Inverted}b) illustrates the mobility of each inverted 2DEG at $\SI{1.3}{K}$ for different electron densities tuned by the back gate. We remove results where parallel conduction sets in corresponding to the $\SI{12}{s}$ and $\SI{18}{s}$ doped samples for densities above $n = 8 \cdot 10^{10}{cm^{-2}}$. The benchmark mobility is achieved by the use of the highest Si concentration ($\SI{18}{s}$), which is also accompanied by the highest electron density. A clear tendency towards lower mobilities is visible from higher to lower doping times, except for the $\SI{3}{s}$ doping. This lowest doped sample even shows almost the identical mobility as the highest doped one, but for lower densities. \\
It is quite intuitive that in the case of very little Si in the doping layer ($\SI{3}{s}$), the 2DEG is less disturbed by RIs compared to other scattering sources, i.e. BIs and IR. An increase of the dopant quantity leads to higher RIs, as is verified in the deteriorated 2DEG quality of the $\SI{6}{s}$ structure. The mobility degradation will continue until a certain doping concentration, we hypothesize that then a smoother potential distribution is created, which in turn reduces RI scattering \cite{friedland1996new,umansky2009mbe}. This is underlined with the $\SI{12}{s}$ and $\SI{18}{s}$ doped structures with enhanced $\mu$ compared to the $\SI{6}{s}$ one. One has also to take into account the higher 2DEG density in case of increased Si doping concentrations. The same reasoning can be applied where the 2DEG starts to screen Coulomb fluctuations coming from the doping layer. \\
We derive the $\alpha$ value from the relation $\mu \propto (n_{2DEG})^{\alpha}$, which is given in the inset of Fig.~\ref{fig:Backgate_Inverted}b). From that, the $\SI{12}{s}$ doped sample is dominated by RI scattering with a relatively large $\alpha = 1.6$. A transition from essentially RI to BI is found for $\SI{18}{s}$ including $\alpha = 0.9$. This perfectly matches with our above described model in the high density regime with RIs becoming less dominant due to a screening interplay between the doping layer and the 2DEG. \\
\newpage

An unusual large exponent around $2$ is found for the $\SI{3}{s}$ and $\SI{6}{s}$ doped samples. In contrast, Saku et al. tuned their completely undoped inverted 2DEGs only by a back gate \cite{saku1998high} resulting in $\alpha \approx 0.65$ by taking their $\mu$ ($n$) results. Interestingly, such high $\alpha$ values were observed earlier in other kind of structures. Conspicuously, they all have a doping layer in between the respective gate and the 2DEG, explaining the difference to the completely undoped 2DEG by Saku et al. \cite{saku1998high}. A further example can be found in the work of Peters et al., where a MODFET and a HIGFET are discussed in combination with a top gate \cite{peters2016gating}. There, an alpha value around $2$ is found only for the MODFET in the low density regime, whereas the HIGFET without any dopants only reveals the usual $\alpha \leq 0.5$. In addition, the high exponent for the MODFET reduces to $1.3$ by tuning the 2DEG towards higher densities. Double layer systems have shown similar high exponents $\alpha = 2.2$. They consist of a back and a top gate combined with a doping layer \cite{scharnetzky2021ion}. Noteworthy is also the similar density range of $\sim n= 0.5 \cdot 10^{11}{cm^{-2}}$ with respect to $\alpha \approx 2$ in case of the MODFET, our back-gated inverted 2DEGs and the double layer structures. \\
Hence, there might be instabilities originating from the doping layer by gate application, which are absent for higher 2DEG densities. In conclusion, a more profound understanding for gated 2DEGs in the presence of a doped layer in particular in the case of low densities is required. Nevertheless our findings are in agreement with our inverted structures without any doping between the interface and the top gate with again $\alpha \approx 0.6$ (Fig.~\ref{fig:Topgate_Inverted}b), chapter~\ref{backgate}). \\ 
In Fig.~\ref{fig:Backgate_Inverted}c), the electron density is given as a function of the back gate voltage. The traces in general follow a linear density increase for increased voltages, except that the $\SI{12}{s}$-and $\SI{18}{s}$ doped structures show a saturation at or above $\SI{-1}{V}$. If we focus on $\sim n = 5 \cdot 10^{10}{cm^{-2}}$ for each sample (cf. circled dots), we can see in Fig.~\ref{fig:Backgate_Inverted}d) that the $\SI{3}{s}$ doped 2DEG exhibits the highest mobility. Remarkably only slightly lower in mobility is the $\SI{18}{s}$ sample followed by the $\SI{12}{s}$\, and $\SI{6}{s}$ ones. This demonstrates the better transport properties of either very high or very low Si concentrations of the $\delta$-doping layer at equal electron densities. \\
In conclusion, it appears that the low Si concentrations are advantageous due to reduced RI scattering, while at (very) high Si concentrations RIs and possibly also BIs become less efficient due to screening. \\

\subsection{Top gating of the inverted heterostructures} \label{topgate}
\begin{figure}[h]
	\centering
	\includegraphics[width=1\textwidth]{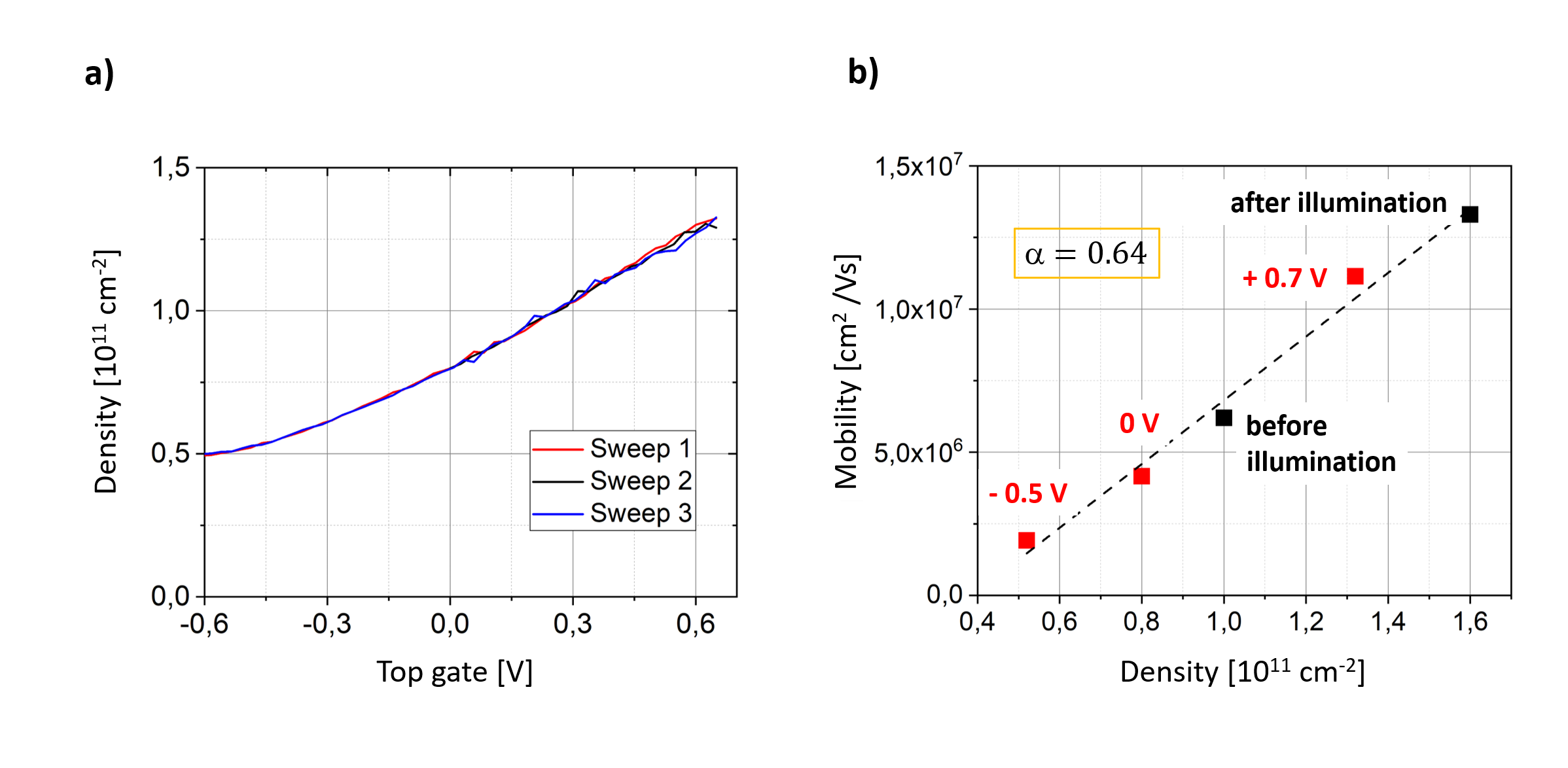}
	\caption{a) Several repetitions of top gate sweeps on a $\SI{1}{\mu m}$ ``deep" inverted 2DEG are shown. Gate voltages from $\SI{-0.6}{V}$ to $\SI{0.6}{V}$ are applied, which results in an electron density change from $n = 0.5 \cdot 10^{11}{cm^{-2}}$ to $n = 1.3 \cdot 10^{11}{cm^{-2}}$. No hysteresis is observed demonstrating a stable gate operation while tuning the 2DEG density. b) A comparison of the mobility as a function of the electron density is illustrated for our benchmark inverted 2DEG once with and once without an Al top gate. Red data points correspond to the gated structure and the black points to the ungated structure, respectively. All data are taken at $\SI{1}{K}$.}
	\label{fig:Topgate_Inverted}
\end{figure}

The opportunity of gate application on a 2DEG provides additional freedom for parameter setting. Besides, it can offer the potential towards hybrid systems where another material type could act also as a gate (see the introduction in chapter~\ref{shallow_inverted}). One important requirement is the gate stability, meaning a suitable gate sweep without any hysteretic effects. \\
For this purpose, we deposited $\sim\SI{100}{nm}$ thick Al films on top of inverted 2DEGs. We use our benchmark inverted 2DEG (Fig.~\ref{fig:Inverted_benchmark_changes}, sample D) with an interface depth of $\SI{1}{\mu m}$\, from the Al layer. In Fig.~\ref{fig:Topgate_Inverted}a), three gate sweeps are shown for a voltage range of $\SI{1.2}{V}$. No hysteresis is observed and the 2DEG density can be varied between $n=0.5 \cdot 10^{11}{cm^{-2}}$ and $n = 1.3 \cdot 10^{11}{cm^{-2}}$. For positive gate voltages (equals higher densities), a linear slope with approx. $5.5 \cdot 10^{10} {cm^{-2}/ V}$ is derived with respect to the capacitance $C = e \cdot \frac{dn}{dV}$ with $e$ the elementary charge. This results in a 2DEG position of almost $d=\SI{1}{\mu m}$, which perfectly matches with the designed value. By comparison, a non-linear behavior of the capacitance occurs for negative voltages (equals lower densities). We attribute this to the appearance of quantum capacitance effects with generally lower total capacitance values \cite{eisenstein1994compressibility}. \\
Furthermore, we compare the top-gated 2DEG with a non-gated reference. Two measurements are given for the reference, meaning one before and one after illumination with the electron density set only by the doped layer in the AlGaAs. These results are summarized in Fig.~\ref{fig:Topgate_Inverted}b). Measurements with the top gate are indicated in red dots with corresponding gate voltages, the reference in black for before and after illumination, respectively. \\
It is quite remarkable that all data points follow the same line, irrespective whether the 2D electrons are obtained via the intrinsic doping layer or via an electric field. This time, the alpha exponent from the relation $\mu \propto (n_{2DEG})^{\alpha}$ is $0.64$ for both examples, well in agreement with our previous inverted 2DEGs described in chapter~\ref{benchmark_inverted}. Moreover, no anomalous enhancement of $\alpha$ is observed as for the back-gated inverted interfaces in chapter~\ref{backgate}. Again the doping layer is not placed between the 2DEG and the top gate. \\

\subsection{40 million mobility in optimized DSQWs} \label{40mio}
\begin{figure}[h]
	\centering
	\includegraphics[width=0.7\textwidth]{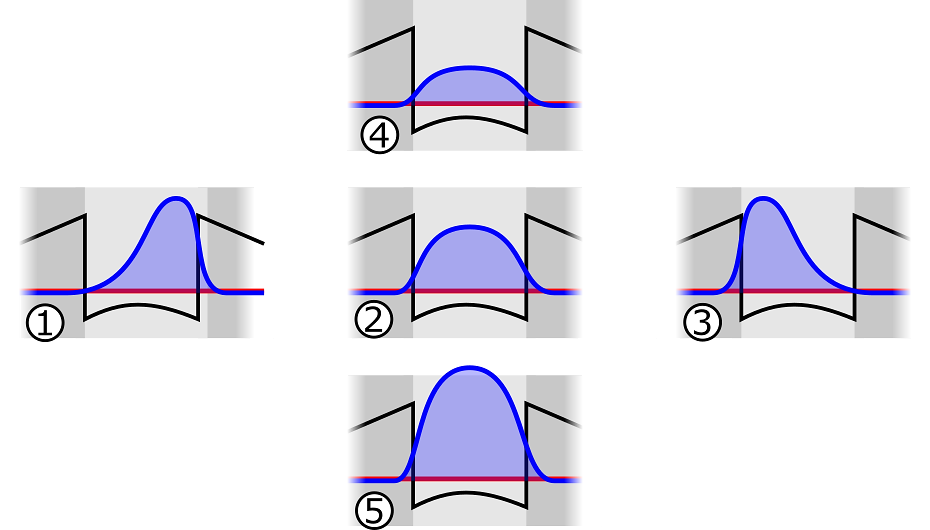}
	\caption{Manipulation the 2DEG`s wave function with separate back and top gate. This allows to push the wave function into each interface ($1$ and $3$) as well as varying the electron density of a centered wave function ($4, 2, 5$).}
	\label{fig:DSQW}
\end{figure}

In this chapter we demonstrate that combining optimized inverted interfaces with top and back gates can yield mobilities of 40 million ${cm^2/Vs}$. These are obtained in double-sided doped quantum well (DSQW) structures, consisting of a central GaAs layer of typically  $\SI{30}{nm}$, enclosed by undoped AlGaAs layers and thin Si doping layers both above and below the quantum well (Fig.~\ref{fig:Structures}d). The two doping layers allow for a higher electron density and thus increased mobility of the 2DEG compared to a single sided doped structure \cite{eisenstein2002insulating}. As the presence of a second interface introduces an additional scattering source it is of particular importance to optimize this (inverted) interface. \\
Using the results described above that led to a significant improvement in the mobility of the inverted 2DEGs, we manage to increase the mobility of our DSQWs from approx. $\mu= 22 \cdot 10^{6}{cm^{2}/Vs}$ to $\mu= 32 \cdot 10^{6}{cm^{2}/Vs}$ (measured at $\SI{1.3}{K}$). Accordingly, main structural modifications are (i) a substrate temperature down-ramping from normally $630 \si{\degree}$C to $480 \si{\degree}$C only during the doping process and (ii) two separate Al concentrations for the buffer and the setback layer (chapter~\ref{benchmark_inverted}). \\
In addition to an optimized inverted interface, we use both a front and a back gate to tune the electron density to an optimal value and position the wave function in between the interfaces to minimize scattering. Back gates for this sample were pre-patterned and then overgrown using the ion-implantation technique developed in our group \cite{berl2016structured}. Separate contacts to the 2DEG and the back gate are produced reliably with this technique (Fig.~\ref{fig:Structures}e). Also the shape of the 2DEG wave function (Fig.~\ref{fig:DSQW}) can be modified by varying the respective gate voltages. We can ``push" the wave function into the inverted and upper interface, respectively, while keeping the electron density in the quantum well constant. Likewise, the density can be varied by over an order of magnitude (from $n=0.4 \cdot 10^{11}{cm^{-2}}$ to $n=4.9 \cdot 10^{11}{cm^{-2}}$), while keeping the wave function centered in the quantum well. This method allows us to study the effects of IR scattering and of electron-electron screening independent from each other. \\

\begin{figure}[h]
	\centering
	\includegraphics[width=1\textwidth]{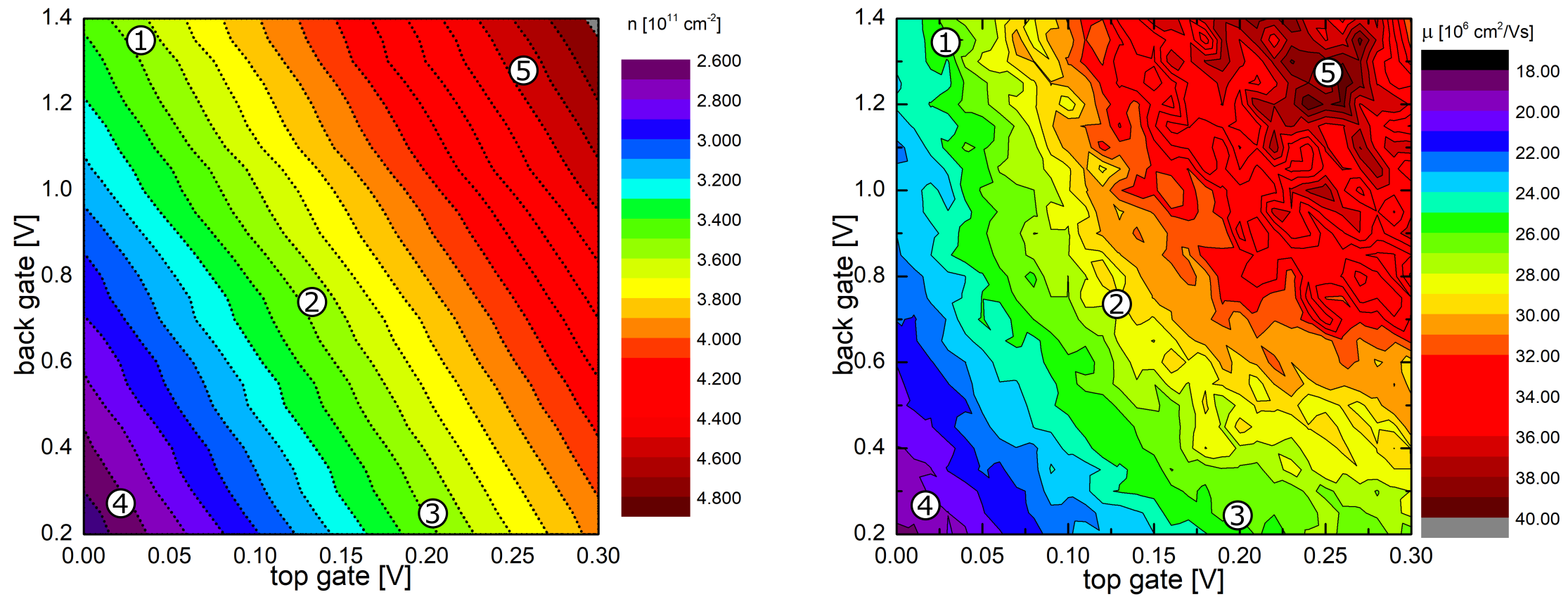}
	\caption{Electron density (left) and mobility (right) mapping, measured at temperatures $< \SI{50}{mK}$,	tuned with a top-and a patterned back gate. Marked with $1, 2, 3$ are gate configurations resulting in identical density, but varying mobility, while $4, 2, 5$ indicate configurations where the 2DEG density is enhanced with a centered wave function.}
	\label{fig:DSQW_Map}
\end{figure}

Fig.~\ref{fig:DSQW_Map} shows the electron density and mobility of the Hallbar device, mapped for back-and top gate voltages tuned separately. Marked with $1, 2, 3$ are gate voltage configurations for maximized scattering at the inverted interface, centered wave function and maximized scattering at the upper interface, yielding electron mobilities of $\mu=25.5 \cdot 10^{6}{cm^{2}/Vs}$, $\mu= 29.0 \cdot 10^{6}{cm^{2}/Vs}$ and $\mu=26.2 \cdot 10^{6}{cm^{2}/Vs}$ respectively. The electron density for each configuration is $n=3.5 \cdot 10^{11}{cm^{-2}}$. The significant, yet very similar, reduction in mobility upon pushing the wave function towards each interface further demonstrates the equal qualities of the AlGaAs/GaAs and the GaAs/AlGaAs interfaces.\\
By varying the density of a centered wave function, indicated by marks $ 4, 2$ and $5$ in Fig.~\ref{fig:DSQW_Map}, the electron mobility changes with density while the impact of IR scattering is minimized. A maximum of mobility is found at mark $5$ measured as $\mu=40 \cdot 10^{6}{cm^{2}/Vs}$ at a density of $n=4.7 \cdot 10^{11}{cm^{-2}}$. (Measurements were performed in an dilution refrigerator at electron temperatures of $< \SI{50}{mK}$.)  \\

\newpage
\section{\textbf{\Large Conclusions and outlook}} 
In summary, we optimized the growth of inverted AlGaAs/GaAs single interface structures towards high low-temperature electron mobilities. Remote ionized impurity (RI) scattering is drastically reduced by employing a substrate temperature down-ramping only during the doping procedure, which limits Si dopant migration along the growth direction. Combination of low and high Al-mol fractions further reduce the effect of background impurities (BIs), remote ionized impurities (RIs) and interface roughness (IR). We achieved a benchmark inverted 2DEG of $\mu = 13 \cdot 10^{6}{cm^{2}/Vs}$ ($ n = 1.6 \cdot 10^{11}{cm^{-2}}$, at $ \SI{1}{K}$). This is of equal quality as the well-established MODFET device confirmed by magnetotransport data. BIs have been determined to remain the dominant limiting factor towards even higher mobilities. \\
Furthermore, we explore RI scattering by Si from a $\delta$-doped AlGaAs layer with respect to the inverted interface quality. Lowest and highest dopant concentrations reveal highest mobilities. The highest doped layer seems to act like a screening layer adding to the intrinsic screening of the 2DEG. Unusual large $\alpha \approx 2$ exponents are derived in the low density regime for low Si concentrations when a global back gate used. Here a more profound scattering analysis is required in particular for different combinations of doping layers and gating. \\
In addition, we discuss growth towards shallow inverted structures with respect to hybrid systems. Samples with interfaces located only $\SI{50}{nm}$ underneath the surface are synthesized showing clear Shubnikov-de Haas oscillations despite rather low mobilities. \\
Top gate stability is excellent in the inverted structures. Further investigations are necessary to decide if the inverted interface structure is superior to the MODFET one for hybrid application. \\
Lastly, our overall understanding enables us to achieve DSQWs of $\mu \approx 40 \cdot 10^{6}{cm^{2}/Vs}$ at $\SI{1}{K}$ by optimizing the location of the electron wave function in the quantum well. In further investigations, this method could be used to further study the individual influences of the interfaces and of the screening mechanisms on the mobility. This capability will be also useful also in the study of exotic properties properties of 2DEG structures of highest quality like fragile fractional quantized Hall states. Particularly, the potentially non-Abelian state at filling factor $\nu = 5/2$ is of special interest, as it is a promising candidate for the realization of topologically protected quantum computing.\\

\section{Acknowledgment}

We gratefully acknowledge the financial support of the Swiss National Foundation (Schweizerischer Nationalfonds, NCCR ``Quantum Science and Technology").

\printbibliography 
\end{document}